\def\be{\begin{equation}}
\def\ee{\end{equation}}
\def\ba{\begin{array}}
\def\ea{\end{array}}
\def\t{\tilde}
\def\L{\Lambda}
\def\l{\lambda}
\def\p{\prime}

\def\Cb{\ \hbox{\vrule width 0.6pt height 6pt depth 0pt
              \hskip -3.2 pt} C}

\documentstyle[12pt]{article}
\begin{document}

\title{ Optimal Teleportation Based on Bell Measurement }
\author{{ Sergio Albeverio$^{a}$
\footnote{E-mail:~albeverio@uni-bonn.de, SFB 256; SFB  237; SFB
611; BiBoS; CERFIM (Locarno); Acc.Arch., USI (Mendrisio)},
Shao-Ming Fei$^{a,b}$ \footnote{E-mail:~fei@uni-bonn.de}, Wen-Li
Yang$^{c,d}$
\footnote{E-mail:~wlyang@th.physik.uni-bonn.de}}\\
{\small $~^{a}$ Institut f\"ur Angewandte Mathematik,
Universit\"at Bonn,
D-53115, Germany}\\
{\small $~^{b}$ Department of Mathematics, Capital Normal University, Beijing
100037, China}\\
{\small $~^{c}$ Physikalisches Institut der Universit\"at Bonn, 53115 Bonn,
Germany}\\
{\small $~^{d}$ Institute of Modern Physics, Northwest University,
Xian 710069, China
} }

\maketitle

\begin{abstract}
We study optimal teleportation based on the Bell measurements. An
explicit expression for the quantum channel associated with the
optimal  teleportation with an arbitrary mixed state resource is
presented. The optimal transmission fidelity of the corresponding
quantum channel is calculated and shown to be related to the fully
entangled fraction of the quantum resource, rather than the singlet
fraction as in the standard teleportation protocol.

\vspace{1truecm}
\end{abstract}


\def\a{\alpha}
\def\b{\beta}
\def\d{\delta}
\def\e{\epsilon}
\def\g{\gamma}
\def\k{\kappa}
\def\l{\lambda}
\def\o{\omega}
\def\p{\phi}
\def\t{\theta}
\def\r{{\cal \rho}}
\def\s{\sigma}
\def\x{\chi}
\def\D{\Delta}
\def\L{\Lambda}
\def\P{\Psi}
\def\Q{\Phi}
\def\F{{\cal F}}
\def\H{{\cal H}}
\def\G{{\cal G}}
\def\Gk{{\cal G}^{(k)}}
\def\R{{\cal R}}
\def\hR{{\hat{\cal R}}}
\def\C{{\bf C}}

\def\uqa{{U_q(\widehat{sl_2})}}
\def\uqA22{{U_q(A^{(2)}_2)}}


\def\beq{\begin{equation}}
\def\eeq{\end{equation}}
\def\bea{\begin{eqnarray}}
\def\eea{\end{eqnarray}}
\def\ba{\begin{array}}
\def\ea{\end{array}}
\def\no{\nonumber}
\def\lt{\left}
\def\rt{\right}
\newcommand{\bq}{\begin{quote}}
\newcommand{\eq}{\end{quote}}

\newtheorem{Theorem}{Theorem}
\newtheorem{Definition}{Definition}
\newtheorem{Proposition}{Proposition}
\newtheorem{Lemma}[Theorem]{Lemma}
\newtheorem{Corollary}[Theorem]{Corollary}
\newcommand{\proof}[1]{{\bf Proof. }
        #1\begin{flushright}$\Box$\end{flushright}}


Quantum teleportation protocols play an important role in quantum
information processing. In terms of a classical communication
channel and a quantum resource (a nonlocal entangled state like an
EPR-pair of particles), the teleportation protocol gives ways to
transmit an unknown quantum state from a sender traditionally
named ``Alice" to a receiver ``Bob" who are spatially separated.
These teleportation processes can be viewed as quantum channels.
The nature of a quantum channel is determined by the particular
protocol and the state used as a teleportation resource
\cite{Ben93, Hor01,Ari00}. The standard teleportation protocol
$T_0$ proposed in \cite{Ben93} uses {\it Bell} measurements and
{\it Pauli} rotations. When the maximally entangled pure state
$|\Q>=\frac{1}{\sqrt{n}}\sum_{i=0}^{n-1}|ii>$ is used as the
quantum resource, it provides an ideal noiseless quantum channel
$\L^{(|\Q><\Q|)}_{T_0}(\r)=\r$. However in realistic situation,
instead of the pure maximally entangled states, Alice and Bob
usually share a mixed entangled state due to the decoherence.
Teleportation using mixed state as an entangled resource is, in
general, equivalent to having a noisy quantum channel. Recently,
an explicit expression for the output state of the quantum channel
associated with the standard teleportation protocol $T_0$ with an
arbitrary  mixed state resource has been obtained
\cite{Bow01,Alb02}.

In this paper, we consider the following problem. Alice and Bob
previously  only share a pair of particles in  an arbitrary mixed
entangled state $\x$. In order to teleport an unknown state to
Bob, Alice first performs a joint Bell measurement on her
particles (particle 1 and particle 2) and tell her result to Bob
by the classical communication channel. Then Bob, instead of the
{\it Pauli} rotation like in the standard teleportation protocol
\cite{Ben93}, tries his best to choose an particular unitary
transformation which depends on the quantum resource $\x$, so as
to get the maximal transmission fidelity. We call our
teleportation protocol the optimal teleportation based on the Bell
measurement. We derive an explicit expression for the quantum
channel associated with the optimal  teleportation with an
arbitrary mixed state resource. The transmission fidelity of the
corresponding quantum channel is given in term of the {\it fully
entangled fraction} of the quantum resource.

Let $\{|i>, i=0,...,n-1\}$, $n<\infty$, be an orthogonal
normalized
basis of an $n$-dimensional Hilbert space $\H$. Any linear
operator $A:~ \H\longrightarrow \H~$ can be represented by an
$n\times n$-matrix as follow: \bea A(|i>)=\sum_{j=0}^{n-1}A_{i
j}|j>,~~~A_{i j}\in\Cb\no \eea We shall only consider the
following three-tensor Hilbert space: $\H\otimes\H\otimes\H$ where
Alice has the first and the second Hilbert space, and the third
one belongs to Bob. Let $h$ and $g$ be $n\times n$ matrices such
that $h|j>=|(j+1)~mod~n>,~~g|j>=\o^j|j>$, with
$\o=exp\{\frac{-2i\pi}{n}\}$. We can introduce  $n^2$
linear-independent $n\times n$-matrices $U_{st}=h^{t}g^s$, which
satisfy \bea
U_{st}U_{s't'}=\o^{st'-ts'}U_{s't'}U_{st},~~tr(U_{st})=n\d_{s0}\d_{t0}.
\label{Mat} \eea One can also check that $\lt\{U_{st}\rt\}$satisfy
the condition of {\it bases of the unitary operators} in the sense
of \cite{Wer00}, i.e.
\bea \lt\{\begin{array}{l}
tr \lt(U_{st}U^+_{s't'}\rt)=n \d_{tt'}\d_{ss'},\\
U_{st}~U_{st}^+=I_{n\times n},
\end{array}\rt.\label{Wer}
\eea where $I_{n\times n}$ is the $n\times n$ identity matrix.
$\lt\{U_{st}\rt\}$ form a complete basis of $n\times n$-matrices,
namely, for any $n\times n$ matrix $W$, $W$ can be expressed as
\bea W=\frac{1}{n}\sum_{s,t}tr (U_{st}^+W)U_{st}.\label{Dec} \eea
From $\left\{U_{st}\right\}$, we can introduce the generalized
Bell-states, \bea |\Q_{st}>=(1\otimes
U^*_{st})|\Q>=\frac{1}{\sqrt{n}}\sum_{i,j}(U_{st})^*_{ij}|ij>
,~~~{\rm and }~~|\Q_{00}>=|\Q>,\label{Bell} \eea $|\Q_{st}>$ are
all maximally entangled states and form a complete orthogonal
normalized basis  of $\H\otimes \H$ shared by Alice and Bob.  For
any state $\x$ shared by Alice and Bob, let us introduce the {\it
singlet fraction} \cite{Hor01}: $F=<\Q|\x|\Q>$. In general, all
the maximally entangled pure states are equivalent to $|\Q>$:
$|\Psi_{max}>=1\otimes U|\Q>$, where $U$ is a unitary
transformation. One can define the {\it fully entangled fraction}
\cite{Hor01} of a state $\x$ by \bea \F(\x)=max\lt\{<\Q|(1\otimes
U^+)~\x~(1\otimes U)|\Q>\rt\},~~ {\rm for~ all}~
UU^+=U^+U=I_{n\times n}.\label{MF1} \eea Since the group of
unitary transformations in n-dimensions is compact, there exists
an unitary matrix $W_{\x}$ such that \bea \F(\x)=<\Q|(1\otimes
W_{\x}^+)~\x~(1\otimes W_{\x})|\Q>.\label{MF2} \eea

Suppose now Alice and Bob  previously share a pair of particles in
an arbitrary mixed entangled state $\x$. To transform  an unknown
state  to Bob, Alice first performs a joint Bell measurement based
on the generalized Bell-states Eq.(\ref{Bell}) on her parties.
According to the measurement results of Alice, Bob chooses
particular unitary transformations $\lt\{T_{st}\rt\}$ to act on
his particle.
\begin{Theorem}
The teleportation protocol defined by $\lt\{T_{st}\rt\}$, when
used with an arbitrary mixed state with density matrix $\x$ as a
resource,  acts as a quantum channel \bea
\L^{(\x)}(\{T\})~(\r)=\frac{1}{n^2}\sum_{s,t}\sum_{s',t'}
<\Q_{st}|\x|\Q_{s't'}>\lt\{\sum_{\g\b}
T_{\g\b}^+U_{st}U_{\g\b}~\r~
U^+_{\g\b}U_{s't'}^+T_{\g\b}\rt\}.\label{Out} \eea
\end{Theorem}

\noindent{\bf Proof}. The proof can be given in two steps:

\noindent{\it Step 1. Pure entangled state as a resource}. Each
entangled pure state $|\P>$ shared by Alice and Bob has the form
\bea |\P>=\sum_{i,j=0}^{n-1}a_{i j}|i j>, ~~
\sum_{i,j=0}^{n-1}|a_{i j}|^2=1,~~~a_{i j}\in\Cb,\label{Ent} \eea
Let $A$ be the $n\times n$ matrix with elements $(A)_{i j}=a_{i
j}$, $a_{i j}\in\Cb$. Suppose Alice wishes to teleport the unknown
pure state $|\p>=\sum_{i=1}^{n}\a_i|i>$. The initial state Alice
and Bob have is then given by \bea |\p>\otimes
|\P>=\sum_{i,j,k=0}^{n-1}\a_ia_{j k}|ij k>~\in \H\otimes \H\otimes
\H.\label{Sta} \eea To transform the state $|\p>$ to Bob, Alice
first performs a joint Bell measurement based on the generalized
Bell-states Eq.(\ref{Bell}) on her party. After her measurement
with outcoming in the state $|\Q_{st}>$, Bob's particle gets into
an (unnormalized) state \bea |\p>\longrightarrow
\frac{1}{\sqrt{n}}AU_{st}|\p>.\no \eea Once Bob learns from Alice
that she has obtained the result $st$, he performs on his
previously entangled particle (particle 3) a unitary
transformation $T_{st}$. Then the final state becomes
$\frac{1}{\sqrt{n}}T^+_{st}AU_{st}|\p>$. In terms of the density
matrix, the teleportation based on the unitary matrices $\left\{
T_{st}\right\}$ , with quantum resource being a pure state $|\P>$,
is a quantum channel with the output \bea
\L^{(|\P><\P|)}(\{T\})~(\r)=\frac{1}{n}\sum_{st}T^+_{st}AU_{st}~\r~
U^+_{st}AT_{st}.\no \eea

\noindent{\it Step 2. An arbitrary mixed entangled state as a
resource.} Let $\x$ be a mixed state, \bea
\x=\sum_{\a}p_{\a}|\P_{\a}><\P_{\a}|,~~0\leq p_{\a}\leq 1 ~~{\rm
and
}~~\sum_{\a}p_{\a}=1,~~|\P_{\a}>=\sum_{i,j}a^{(\a)}_{ij}|ij>.\no
\eea Applying the teleportation protocol $T$ with a mixed state
$\x$, the final state of Bob becomes \bea
\L^{(\x)}(\{T\})~(\r)=\frac{1}{n}\sum_{s,t}\sum_{\a}p_{\a}T^{+}_{st}A^{(\a)}U_{st}~\r~
U^{+}_{st}(A^{(\a)})^+T_{st}.\label{Tel} \eea Since each matrix
$A^{(\a)}$ can be decomposed in the basis of $\lt\{U_{st}\rt\}$ by
$(A^{(\a)})_{ij}=\sum_{s,t}a^{(\a)}_{st}(U_{st})_{ij}$,
(\ref{Tel}) becomes \bea
\L^{(\x)}(\{T\})~(\r)&=&\frac{1}{n}\sum_{s,t}\sum_{s',t'}\lt(
\sum_{\a}p_{\a}a^{(\a)}_{st}a^{(\a)*}_{s',t'}\rt)\sum_{\g,\b}
T_{\g\b}^+U_{st}U_{\g\b}~\r~ U^+_{\g\b}U_{s't'}^+T_{\g\b}.\no \eea
Using the definition of generalized Bell-states
$\lt\{|\Q_{st}>\rt\}$ in Eq.(\ref{Bell}), after a lenthy
calculation, we arrive at \bea
n\sum_{\a}p_{\a}A^{(\a)}_{st}A^{(\a)*}_{s',t'}=<\Q_{st}|\x|\Q_{s't'}>.\no
\eea Substituting the above results into Eq.(\ref{Tel}), one
obtains Eq.(\ref{Out}). Using Eq.(\ref{Wer}) and the identity \bea
\sum_{s,t}U^+_{st}~A~U_{st}=n\, tr(A)\, I_{n\times n},~~{\rm for
~any~} n\times n ~{\rm matrix}~ A,\no \eea the trace-preserving
property of  the quantum channel can be proved by \bea
tr\lt(\L^{(\x)}(\{T\})~(\r)\rt)
&=&\frac{1}{n^2}\sum_{s,t}\sum_{s',t'}
<\Q_{st}|\x|\Q_{s't'}>\lt\{\sum_{\g\b}~tr\lt(
U^+_{\g\b}U_{s't'}^+U_{st}U_{\g\b}\r \rt)\rt\}\no\\
&=&\frac{1}{n}\sum_{s,t}\sum_{s',t'}
<\Q_{st}|\x|\Q_{s't'}>tr\lt(U_{st}U_{s't'}^+\rt)\times tr(\r)\no\\
&=&\sum_{s,t}<\Q_{st}|\x|\Q_{st}>=tr(\x)=1.\no
\eea
$\Box$

The fidelity of the teleportation is given by \bea
f(\x)=\overline{<\p_{in}|
\L^{(\x)}(\{T\})~(|\p_{in}><\p_{in}|)|\p_{in}>},\label{Fed} \eea
averaged over all pure input states $\p_{in}$.

In order to calculate the transmission fidelity Eq.(\ref{Fed}), we
need an irreducible n-dimensional representation of the unitary group
${\bf
U(n)}$, denoted by {\bf G}. Let $U(g)$ be the unitary matrix
representation of the element $g$ of {\bf G}. Recalling Schur's
Lemma, one has the identity \bea &&\int_{{\bf G}}
dg~(U^+(g)\otimes U^+(g))~\s~(U(g)\otimes U(g))=\a_1I\otimes
I+\a_2 P,
\label{Shu}\\
&&\a_1=\frac{n^2tr(\s)-ntr(\s P)}{n^2(n^2-1)},~~
\a_2=\frac{n^2tr(\s P)-ntr(\s )}{n^2(n^2-1)},\no \eea for any
operator $\s$ acting on the tensor space, where $P$ is the flip
operator such that $P|ij>=|ji>$. The invariant (Haar) measure $dg$
on $G$ is normalized by $\int_{{\bf G}}dg=1$.

\begin{Theorem}
The transmission fidelity of the teleportation protocol defined by
$\lt\{T_{st}\rt\}$ with arbitrary mixed state $\x$ as a resource
is given by \bea f(\x)=\frac{1}{n(n+1)}\sum_{\g\b}<\Q|\lt(1\otimes
(T_{\g\b}U^{+}_{\g\b})^+\rt)~\x~ \lt(1\otimes
T_{\g\b}U^+_{\g\b}\rt)|\Q>~+\frac{1}{n+1}.\label{Fe} \eea
\end{Theorem}
\vspace{0.4truecm}
\noindent{\bf Proof}. From Theorem 1 and Eq.(\ref{Shu}), one has
\bea
f(\x)
&=&\frac{1}{n^2}\sum_{s,t}\sum_{s',t'}<\Q_{st}|\x|\Q_{s't'}>
\sum_{\g\b}\overline{<\p_{in}|
T^+_{\g\b}U_{st}U_{\g\b}|\p_{in}><\p_{in}|U^+_{\g\b}U^+_{s't'}T_{\g\b}
|\p_{in}>}\no\\
&=&\frac{1}{n^2}\sum_{s,t}\sum_{s',t'}<\Q_{st}|\x|\Q_{s't'}>
\sum_{\g\b}\overline{<\p_{in}|\otimes<\p_{in}|\lt(
T^+_{\g\b}U_{st}U_{\g\b}\otimes U^+_{\g\b}U^+_{s't'}T_{\g\b}\rt)
|\p_{in}>\otimes|\p_{in}>}\no\\
&=&\frac{1}{n^2}\sum_{s,t}\sum_{s',t'}<\Q_{st}|\x|\Q_{s't'}>
\sum_{\g\b}<00| \int_{{\bf G}} dg~(U(g)^+\otimes U(g)^+)\no\\
&&~~~~~~~~~~~~~~~~~~~~~~~~~~~~~~~~\times \lt(
T^+_{\g\b}U_{st}U_{\g\b}\otimes U^+_{\g\b}U^+_{s't'}T_{\g\b}\rt)(U(g)\otimes U(g))
|00>\no\\
&=&\frac{1}{n^3(n+1)}\sum_{s,t}\sum_{s',t'}<\Q_{st}|\x|\Q_{s't'}>
\sum_{\g\b}\lt\{
tr\lt(T^+_{\g\b}U_{st}U_{\g\b}\rt)\times~tr\lt(
U^+_{\g\b}U^+_{s't'}T_{\g\b}\rt)\rt.\no\\
&&~~~~~~~~~~~~~~~~~~~~~~~~~~~~~~~~~~+\lt.tr\lt(T^+_{\g\b}U_{st}U_{\g\b}
U^+_{\g\b}U^+_{s't'}T_{\g\b}\rt)\rt\}\no\\
&=&\frac{1}{n(n+1)}\sum_{\g\b}<\Q|\lt(1\otimes
(T_{\g\b}U^{+}_{\g\b})^+\rt)~\x~ \lt(1\otimes
T_{\g\b}U^+_{\g\b}\rt)|\Q>~+\frac{1}{n+1}.\no \eea where  the
identity $tr_{12}\lt((A\otimes B)P\rt)=tr(AB)$, Eq.(\ref{Wer}) and
Eq.(\ref{Dec}) have been used. $\Box$

Obviously when the term $<\Q|\lt(1\otimes
(T_{\g\b}U^{+}_{\g\b})^+\rt)~\x~ \lt(1\otimes
T_{\g\b}U^+_{\g\b}\rt)|\Q>$ is maximized, i.e.,
$T_{\g\b}U^{+}_{\g\b}=W_\x$, one gets the maximal fidelity.
Recalling the definition of the {\it fully entangled fraction}
Eq.(\ref{MF1}) and Eq.(\ref{MF2}), we arrive at our main result:
\begin{Theorem}
The optimal teleportation based on the Bell measurements, when
used with an arbitrary mixed state with density matrix $\x$ as a
resource, acts as a general trace-preserving quantum channel \bea
\L^{(\x)}_{O}~(\r)=\frac{1}{n^2}\sum_{s,t}\sum_{s',t'}
<\Q_{st}|\x|\Q_{s't'}>\lt\{\sum_{\g\b}
U_{\g\b}^+W_{\x}^+U_{st}U_{\g\b}~\r~
U^+_{\g\b}U_{s't'}^+W_{\x}U_{\g\b}\rt\}. \eea The corresponding
transmission fidelity is given by \bea
f_{max}(\x)=\frac{n\F(\x)}{n+1}+\frac{1}{n+1}, \eea where $\F(\x)$
is the fully entangled fraction Eq.(\ref{MF1})and $W_{\x}$ is the
unitary matrix which fulfills such a fully entangled fraction
Eq.(\ref{MF2}).
\end{Theorem}

Our results show  that  the maximally transmission fidelity of the
teleportation based on the Bell measurement depends on the {\it
fully entangled fraction} only, whereas that of a standard
teleportation depends on the singlet fraction \cite{Alb02}.  Our
result also agrees with the fidelity formula of the general
optimal teleportation given by the Horodecki family \cite{Hor99}.

Summarizing, we obtain  the explicit expression of the output
state of the optimal teleportation, with arbitrary mixed entangled
state as resource,  in terms of some noisy quantum channel. This
allow us to calculate the transmission fidelity of the quantum
channel. It is shown that the transmission fidelity depends only
on  the {\it fully entangled fraction} of the quantum resource
shared by the sender and the receiver. The fidelity in our optimal
teleportation protocol is in general greater than the one in
standard teleportation protocol \cite{Ben93,Bow01,Alb02}.

\vspace{1.0truecm}

\noindent {\bf Acknowledgments.} W.-L. Yang would like to thank
Prof. von Gehlen for his continuous encouragements. He has been
supported by the Alexander-von-Humboldt Foundation.
\smallskip

\bibliographystyle{unsrt}

\end{document}